\documentclass[pra,twocolumn,superscriptaddress,nofootinbib]{revtex4-1}
\usepackage{graphicx}
\usepackage{bm}
\usepackage{amssymb,amsmath,mathrsfs}
\usepackage{color}
\usepackage{amstext}
\usepackage[english]{babel}
\usepackage{latexsym}
\usepackage{array}   
\usepackage{multirow}         
\usepackage[usenames,dvipsnames]{xcolor}
\usepackage[colorlinks=true,citecolor=Blue,linkcolor=RubineRed,urlcolor=Blue]{hyperref}
\usepackage[all]{xy}     
\usepackage[usenames,dvipsnames]{xcolor}

\definecolor{mygreen}{RGB}{0,110,62}
\definecolor{mypurple}{RGB}{134,3,191}

\newcommand{\pac}[1]{ \left\{ #1 \right\} }
\newcommand{\pap}[1]{\left( #1 \right)}
\newcommand{\pas}[1]{\left[#1 \right]}

\newcommand{\ket}[1]{ \left| #1 \rangle\right.}
\newcommand{\bra}[1]{  \left.\langle #1  \right|}

\newcommand{\mygreen}[1]{{\color{mygreen} #1}}
\newcommand{\mypurple}[1]{{\color{mypurple} #1}}

\begin{document}

\title{Exploring Supersymmetry: Interchangeability Between Jaynes-Cummings and Anti-Jaynes-Cummings Models}

\author{Ivan~A. Bocanegra-Garay\href{https://orcid.org/0000-0002-5401-7778}{\includegraphics[scale=0.45]{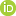}}}
\email[]{ivanalejandro.bocanegra@uva.es}
\affiliation{Departamento de F\'isica Te\'orica, At\'omica y \'Optica and  Laboratory for Disruptive Interdisciplinary Science, Universidad de Valladolid, 47011 Valladolid, Spain}

\author{Miguel Castillo-Celeita\href{https://orcid.org/0000-0003-2905-0389}{\includegraphics[scale=0.45]{orcid}}}
\affiliation{Departamento de F\'isica Te\'orica, At\'omica y \'Optica and  Laboratory for Disruptive Interdisciplinary Science, Universidad de Valladolid, 47011 Valladolid, Spain}

\author{J.~Negro\href{https://orcid.org/0000-0002-0847-6420}{\includegraphics[scale=0.45]{orcid}}}
\affiliation{Departamento de F\'isica Te\'orica, At\'omica y \'Optica and  Laboratory for Disruptive Interdisciplinary Science, Universidad de Valladolid, 47011 Valladolid, Spain}

\author{ L.~M. Nieto\href{https://orcid.org/0000-0002-2849-2647}{\includegraphics[scale=0.45]{orcid}}}
\affiliation{Departamento de F\'isica Te\'orica, At\'omica y \'Optica and  Laboratory for Disruptive Interdisciplinary Science, Universidad de Valladolid, 47011 Valladolid, Spain}

\author{Fernando~J. G\'omez-Ruiz\href{https://orcid.org/0000-0002-1855-0671}{\includegraphics[scale=0.45]{orcid}}}
\email[]{fernandojavier.gomez@uva.es}
\affiliation{Departamento de F\'isica Te\'orica, At\'omica y \'Optica and  Laboratory for Disruptive Interdisciplinary Science, Universidad de Valladolid, 47011 Valladolid, Spain}

\date{\today}

\begin{abstract}
The supersymmetric connection that exists between the Jaynes-Cummings (JC) and anti-Jaynes Cummings (AJC) models in quantum optics is unraveled entirely.
A new method is proposed to obtain the temporal evolution of observables in the AJC model using supersymmetric techniques, providing an overview of its dynamics and extending the calculation to full photon counting statistics. The approach is general and can be applied to determine the high-order cumulants given an initial state. The analysis reveals that engineering the collapse-revival behavior and the quantum properties of the interacting field is possible by controlling the initial state of the atomic subsystem and the corresponding atomic frequency in the AJC model. The substantial potential for applications of supersymmetric techniques in the context of photonic quantum technologies is thus demonstrated.
\end{abstract}
\maketitle

\section*{Introduction}
\begin{figure}[t!]
\centering
\includegraphics[width=1\linewidth]{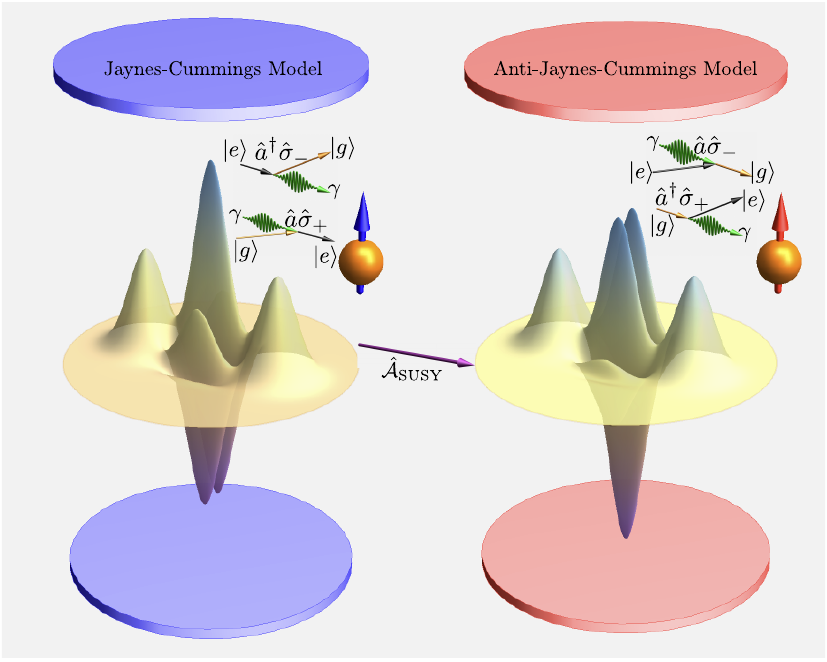}
\caption{\label{fig_1}  {\bf Schematic representation of SUSY mapping in quantum optics models.} Experimental setups for studying the JC and AJC models traditionally involve confining a single electromagnetic mode between two mirrors. This mode interacts with an atomic system possessing two energy levels, $\{\ket{g},\ket{e}\}$. In the JC model Eq.~\eqref{JCM_Ham}, excitations can be exchanged between photons ($\gamma$) and atomic energy levels using operators such as $\hat{a}^{\dagger}\hat{\sigma}_{-}$ and $\hat{a}\hat{\sigma}_{+}$. On the contrary, the AJC model Eq.~\eqref{AJCM_Ham} allows for simultaneous creation or annihilation processes in both subsystems, as described by operators $\hat{a}^{\dagger}\hat{\sigma}_{+}$ and $\hat{a}\hat{\sigma}_{-}$. }
\end{figure}

The introduction of the Jaynes-Cummings (JC) model in 1963~\cite{JC_Model63} marked a crucial moment in the history of quantum optics and continues to be a cornerstone of research in this discipline today, as it captures the fundamental dynamics of the radiation-matter interaction within the weak coupling quantum regime.
This model relies on three key approximations: (i) the reduction of matter-light interaction to the dipolar regime, (ii) the portrayal of subsystem matter as a quantum two-level system, and (iii) the conservation of total excitations, often referred to as the { \it``rotating wave approximation''}. Consequently, the traditional form of the Hamiltonian for the JC model is expressed as follows (from now on $\hbar =1$)
\begin{equation}\label{JCM_Ham}
 \hat{H}_{\rm JC} 
 =\frac{\omega_a}{2}\hat{\sigma}_z + \omega_c \hat{a}^{\dagger}\hat{a}+\lambda\pap{\hat{a}\hat{\sigma}_{+}+\hat{a}^{\dagger}\hat{\sigma}_{-}},
\end{equation}
where the light-matter coupling is given by $\lambda$, representing $\omega_c$ and $\omega_a$, respectively, the fundamental energies of the single-mode quantized electromagnetic field in the cavity and of the two-level atomic system. The matter operators can be expressed in the basis of the excited state $\ket{e}$ and the ground state $\ket{g}$ of the atom as $\hat{\sigma}_z=\vert e\rangle\langle e\vert -\vert g\rangle\langle g\vert$, $\hat{\sigma}_{-}=\vert g\rangle\langle e\vert$, and $\hat{\sigma}_{+}=\vert e\rangle\langle g\vert$. Additionally, the operators $\hat{a}^{\dagger}$ and $\hat{a}$ represent the traditional creation and annihilation photon operators acting on the usual Fock or number states $\ket{n}$. The action of these operators that make up $\hat{H}_{\rm JC}$ is schematically depicted on the left side of Fig~\ref{fig_1}.
Over the past few decades, rapid advancements in quantum technologies have enabled the experimental realization of the JC model in laboratory settings. This versatile model has found applications across a spectrum of physical platforms. These include neutral atoms confined within optical and microwave cavities~\cite{Brune_PRL96}, trapped ions~\cite{Diedrich_PRL89,Blockley_ELet92}, and superconducting qubits coupled with electromagnetic cavities~\cite{Blais_RModPhys21}, transmission line resonators~\cite{Wallraff_Nat04}, and cold atoms~\cite{Ritsch_RModPhys13}, among others. For a comprehensive review, please refer to Ref.~\cite{Larson_2021}.

Additionally, due to the success of the JC model, several researchers have begun to explore a counterpart version of it. This alternative version, known as the anti-Jaynes-Cummings (AJC) model~\cite{Solano_PRL03, Bermudez_PRA07}, modifies the radiation-matter interaction term by allowing only those terms that do not conserve the number of excitations, as schematically shown in the right side of Fig.~\ref{fig_1}. This characteristic sets it apart from the JC model and opens up new avenues for exploring non-conservative dynamics in quantum systems. Therefore, the Hamiltonian takes the following form:
\begin{equation}\label{AJCM_Ham}
 \hat{H}_{\rm AJC} 
=\frac{\omega_a}{2}\hat{\sigma}_z + \omega_c \hat{a}^{\dagger}\hat{a}+\lambda\pap{\hat{a}^{\dagger}\hat{\sigma}_{+}+\hat{a}\hat{\sigma}_{-}}.
\end{equation}
The exploration of the AJC model has revealed intriguing properties, drawing parallels with the JC model~\cite{Lara_PRA05, Uhdre_PRA22, Mayero2023, Choreo2017, Raffa2017}, thereby providing fresh perspectives and insights that deepen our understanding of quantum dynamics.

In a parallel historical context, supersymmetry (SUSY) has fascinated particle physicists since its inception~\cite{Cooper1995}. Notably, seminal ideas have brought SUSY concepts into the realm of quantum optics, particularly within the JC model~\cite{Andreev_PLA89, Hussin_2006, Fan_1995}. Recently, Rodr\'iguez-Lara and collaborators have conducted significant work on exploring the relationship between quantum optics models and SUSY~\cite{Maldonado2021, Rodriguez-Lara:14,kafuri2024}. Despite these advances, an important gap remains a clear connection between two of the most paradigmatic models in quantum optics, the JC model, and its counterpart, the AJC model.

In this work, we address this gap by presenting a new approach to map the JC model solutions into the AJC model ones. We introduce a SUSY transformation that allows a clear mapping between the JC and AJC Hamiltonians. This work presents extensive theoretical and numerical studies of the SUSY connection between both models. Additionally, it provides a general conceptual framework for obtaining high-order cumulants in the context of full counting statistics of photons.

The paper is organized as follows. In Sec.~\ref{Teo_Back}, we recapitulate the theoretical background of the JC model for the exact dynamical solution. In Sec.~\ref{Sec_SUSY_I}, we introduce the SUSY mapping between the JC and AJC models. Specifically, we define a  SUSY transformation that allows us to map the time-dependent solutions of the JC into the AJC model. We demonstrate that the partner Hamiltonian of the JC model is the AJC model with a shifted qubit frequency. In Sec.~\ref{Sec_Exact_Dyn}, we present a comprehensive analysis of the expectation values of qubit and field operators. For the field operators, we extend the calculation to obtain key properties of the counting statistics of photons. In Sec.~\ref{Sec_Num_Exp}, we present numerical experiments demonstrating sensible results regarding the control of collapse-revival behavior and the quantum nature of the radiation in the AJC model. The work is closed with a final section of conclusions, where some interesting possibilities of future work are indicated, for the study of more sophisticated systems. A couple of  Appendices are given for the sake of completeness.
\section{Theoretical background}\label{Teo_Back}

As mentioned in the introduction, the JC model has been one of the most studied from both theoretical and experimental perspectives. In this section, we provide a brief overview of the solution of the corresponding Schr\"odinger equation, through the evolution operator for the JC model. These results will be employed throughout the manuscript, aiming to create a self-contained work. The Hamiltonian of the  JC model, given in Eq.~\eqref{JCM_Ham}, can be written in matrix form as:
\begin{equation}\label{hoperator}
  \hat{H}_{\rm JC}
    = \begin{pmatrix}
    \omega_c \hat{n} +\frac{\omega_a}{2}\hat{\mathbb{I}} & \lambda\hat{a} \\
    \lambda\hat{a}^\dagger & \omega_c \hat{n} -\frac{\omega_a}{2}\hat{\mathbb{I}}
    \end{pmatrix},
\end{equation}
where $\hat{n}=\hat{a}^{\dagger}\hat{a}$ is the operator representing the number of photons and $\hat{\mathbb{I}}$ represents the identity operator in Fock space. 
In general, the solution of the Schr\"odinger equation $i\partial_t \ket{\psi_{\rm JC}(t)} = \hat{ H}_{\rm JC} \ket{\psi_{\rm JC}(t)}$ is given by $\ket{\psi_{\rm JC}(t)}=\hat{U}(t)\ket{\psi_{\rm JC}(0)}$, 
where $\ket{\psi_{\rm JC}(0)}$ is the initial state of the total system: light and matter. As the Hamiltonian (Eq.~\eqref{hoperator}) is time-independent, the unitary evolution operator $\hat{U}(t)$ can be expressed as~\cite{Gerry2004}:  
\begin{equation}\label{U_I}
\hat{U}\pap{t}=\exp\pas{-it\hat{H}_{\rm JC} }
=\begin{pmatrix} 
\mathcal{F}_{\hat{n}+\hat{\mathbb{I}}}^{\dagger}(t) & \mathcal{G}_{\hat{n},\hat{a}}\pap{t} \\[1ex] 
\mathcal{G}_{\hat{n}+\hat{\mathbb{I}},\hat{a}^{\dagger}}\pap{t}  & \mathcal{F}_{\hat{n}}(t)  
\end{pmatrix},
\end{equation}
where the matrix elements are given by the three nontrivial operators
\begin{equation}
\begin{split}\label{func_UI}
\mathcal{F}_{\hat{m}}&=\cos \pas{\Omega_{\hat{m}} t} + i \,\frac{\Delta}{2}
\;\frac{\sin \pas{\Omega_{\hat{m}} t }}{\Omega_{\hat{m}}},
\\
\mathcal{G}_{\hat{m},\hat{O}}&=-i\, \lambda\;\hat{O}\;\frac{\sin \pas{\Omega_{\hat{m}} t }}{\Omega_{\hat{m}}},
\\
\Omega_{\hat{m}} &= \sqrt{\pap{\Delta/2}^2\hat{\mathbb{I}} + \lambda^2 \hat{m}},
\end{split}
\end{equation}
$\hat{O}$ being any of the three operators $\hat{\mathbb{I}}$, $\hat{a}$ or $\hat{a}^\dagger$, and
the photon index operator $\hat{m }$ being either $\hat{n}$ or $\hat{n} + \hat{\mathbb{I}}$.
Here we are denoting as $\Omega_{\hat{m}}$ the Rabi-modified frequency operator and by $\Delta = \omega_a - \omega_c$ the detuning between the atomic and field frequencies. 
In particular, we can give specific cases of the expected value of functions defined by Eq.~\eqref{func_UI} as 
\begin{subequations}
\begin{align}
\mathcal{F}_{m}& =\bra{m}\mathcal{F}_{\hat{m}}\ket{m} =\cos (\Omega_{m} t) + i \frac{\Delta}{2} \frac{\sin (\Omega_{m} t)}{ \Omega_{m}},
\label{Exp_F}
\\
\mathcal{G}_{m}& =\bra{m}\mathcal{G}_{\hat{m},\hat{I}}\ket{m} =-i \lambda\ \frac{\sin (\Omega_{m} t )}{\Omega_{m}},
\label{Exp_G}
\end{align}
\end{subequations}
with $\Omega_{m}= \sqrt{\pap{\Delta/2}^2  + \lambda^2 m}$.  Note that the functions $\mathcal{F}_{m}$ and $\mathcal{G}_{m}$ are time-dependent. A rather general initial state of the JC model at $t=0$ can be written as: 
\begin{equation}\label{gic}
\ket{\psi_{\rm JC}(0)} =\frac{1}{\mathcal{N}_0}(
\beta_{e}\ket{e}+
\beta_{g}\ket{g})\otimes \sum_{n=0}^{\infty}C_{n}\ket{n},
\end{equation}
where the field probability amplitudes $C_{n}$ satisfy $\sum_{n=0}^{\infty}\vert C_{n}\vert^2 = 1$, and $\mathcal{N}_0 = \sqrt{\vert\beta_{g}\vert^2+\vert\beta_{e}\vert^2}$ is the normalization constant. 
The parameters $\beta_{i} \in \mathbb{C}$, with $i\in\pac{e,g}$, represent the corresponding initial probability amplitudes of finding the atomic system either in the excited state $\ket{e}$ or in the ground state $\ket{g}$, respectively.
Note that we are considering a state of light written as a superposition of Fock states $\ket{n}$ with complex amplitudes $C_{n}$. Using the exact analytical results provided by Eq.~\eqref{U_I} and Eq.~\eqref{func_UI}, the evolution operator is to be applied to any initial state (Eq.~\eqref{gic}) to obtain the evolved wave function $\ket{\psi_{\rm JC}(t)} =\hat{U}\pap{t} \ket{\psi_{\rm JC}(0)}$.  

The dynamics of the JC model captures the fundamental interaction between a two-level atom and a single-mode quantized electromagnetic field~\cite{Varcoe_Nat00, OConnell_Nat10, BenHayun_SciAdv21}, giving rise to intriguing phenomena such as the collapse and revival in the populations~\cite{Haroche_RMPhys13, Wineland_RMPhys13, Moya3_PRA92, Julio_PR90}. These phenomena, characterized by the periodic disappearance and reappearance of the excited state population~\cite{Haroche_RMPhys13, Wineland_RMPhys13}, have been extensively studied and observed experimentally in various physical systems~\cite{Moya2_PRA93, Bocanegra_SciPost24}. In addition to collapse and revival, the JC model also exhibits other fascinating features such as vacuum Rabi oscillations~\cite{Haroche_PRL83, Haroche_PRL01, Wineland_RMPhys99}, quantum interference~\cite{Moya_PRA93, Moya4_PRA03}, and entanglement generation between the atom and the field~\cite{Haroche_RMPhys01, Dutra_PRA94,  Phoenix_PRA91} (see also Refs.~\cite{BrubLauzire1994,Hussin2005}). These details have aroused significant interest due to their potential applications in quantum information processing~\cite{Karnieli2_SciAdv23, Blais2020, Mischuck_PRA13}, quantum communication~\cite{Lucero_NatPhys12, Haroche_NatPhy2020}, and quantum metrology~\cite{Karnieli_SciAdv23, Haroche_PRA95}. In the next section, we establish a supersymmetric (SUSY) approach that will allow us to map the dynamics of the JC model into that of the AJC model in a straightforward way.

\section{Duality in quantum dynamics:  JC and its  SUSY counterpart, the  AJC Hamiltonian}\label{Sec_SUSY_I}

The SUSY approach introduces the  operator $\hat{\mathcal{A}}$ intertwining~\cite{Samsonov_2004,Hussin_2006,Cooper1995,fernandez1999,Mielnik2000,Mig012014,Mig032019,Mig042023,miri012013,Correa2015,Cen2019,RosasOrtiz2020,CruzyCruz2020,CruzyCruz2021,BocanegraGaray2024} two Hamiltonians $\hat{H}_1$ and $\hat{H}_2$ as
\begin{equation}\label{intert}
\hat{\mathcal{A}} \hat{H}_1 = \hat{H}_2 \hat{\mathcal{A}}.
\end{equation}
By choosing $\hat{H}_1=\hat{H}_{\rm JC}$ given in Eq.~\eqref{hoperator}, and taking the operator 
\begin{equation}\label{Loper}
\hat{\mathcal{A}}
 =\begin{pmatrix}
     \hat{a}^\dagger & 0 \\
    0 & \hat{a}
    \end{pmatrix}, 
\end{equation} 
generating a new matrix Hamiltonian $\hat{H}_2$ in the form
\begin{equation}\label{h_SUSY_AJC}
    \hat{H}_2 =\begin{pmatrix}
    \omega_c\hat{n}  +\pap{\frac{\omega_a}{2}-\omega_c}\hat{\mathbb{I}} & \lambda\hat{a}^\dagger\\
    \lambda\hat{a} & \omega_c\hat{n}  - \pap{\frac{\omega_a}{2}-\omega_c}\hat{\mathbb{I}}
    \end{pmatrix},
\end{equation}
which is precisely the AJC Hamiltonian. It is important to note that the atomic frequency in $\hat{H}_2$ is not the same as in $\hat{H}_1$, but it is displaced by the field frequency due to Eq.~\eqref{intert}: the JC Hamiltonian, under an intertwining transformation given by Eq.~\eqref{Loper}, produces an AJC Hamiltonian, which is its SUSY partner, with a modified atomic frequency. Here, we rewrite Eq.~\eqref{intert} as: 
\begin{equation}\label{L_oper_JC}
\hat{\mathcal{A}}\:\hat{H}_{{\rm JC}}\pap{\omega_a}=\hat{H}_{{\rm AJC}}\pap{\omega_a-2\omega_c}\:\hat{\mathcal{A}},
\end{equation}
where, for the sake of clarity, we have now explicitly stated the value of the atomic frequency on which each Hamiltonian depends. However, to keep the notation as simple as possible, the change in atomic frequency in Eq.~\eqref{L_oper_JC} will not be shown explicitly from now on. In Appendix~\ref{apnedix_01}, some interesting properties derived from Eq.~\eqref{L_oper_JC} are presented.

If we assume that $f(z)$ admits a Taylor series expansion, then it makes sense to consider $f(\hat{H})$, and we have that the operator relation given by Eq.~\eqref{L_oper_JC} can be written in a more general form as:
\begin{equation}\label{gintert}
    \hat{\mathcal{A}} f\pap{\hat{H}_{\rm JC}} =f\pap{\hat{H}_{\rm AJC}}\hat{\mathcal{A}}.
\end{equation}
In Section~\ref{Teo_Back} we have presented the exact analytical solution of the JC model, from which the corresponding observables and the expected values of the relevant dynamical variables can be easily obtained. In the following, we will develop a novel approach to obtain the solution of the AJC model by starting from the solution $\ket{\psi_{\rm JC}(t)}$ of the JC model. In turn, this will provide a direct way to obtain the observables and expectation values of dynamical variables associated with the AJC model.

\section{Exact quantum dynamics unraveled through SUSY mapping}\label{Sec_Exact_Dyn}

\begin{figure}[t!]
\centerline{
\xymatrixrowsep{4pc}\xymatrixcolsep{5pc}
\xymatrix@-160pc@R=70pt@C=100pt{
\mygreen{\ket{\psi_{\rm JC}(0)}} \ar@[red][r] \ar@{=>}[r]^{\hat{U}_{\rm JC}\pap{t}} \ar@[blue][d]\ar@{=>}[d]_{\hat{\mathcal{A}}} & \mygreen{\ket{\psi_{\rm JC}(t)}} \ar@[red][d]\ar@{=>}[d]^{\hat{\mathcal{A}}} \\ \mypurple{\ket{\phi_{\rm AJC}(0)}} \ar@[blue][r]\ar@{=>}[r]_{\hat{U}_{\rm AJC}\pap{t}} & \mypurple{\ket{\phi_{\rm AJC}(t)}}
}
}
\caption{\label{Diag_Map}{\bf Schematic Commutative Diagram for Mapping JC Solutions into the AJC Solutions.} The time-dependent solution of the AJC model can be obtained from the evolved solution corresponding to the JC model. This is represented by the path of red arrows (right-down), where $\hat U_{\rm JC}(t)$ is the evolution operator given in Eq.~\eqref{U_I}. Alternatively, the solution of the AJC model can be obtained by applying the intertwining operator to the initial state of the JC model and then performing the time evolution through $\hat{U}_{\rm AJC}\pap{t} = \exp\pas{-it\hat{H}_{\rm AJC} }$. This in turn is depicted by the path of blue arrows (down-right).}
\end{figure}
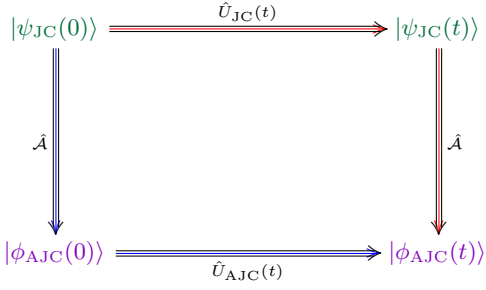

In this section, we present a comprehensive approach to mapping the exact dynamics of the  JC into that of the AJC model. The AJC  dynamics is characterized by the state $ \ket{\phi_{\rm AJC}(t)}$, which is the solution of the Schr\"odinger equation $  i\partial_t \ket{\phi_{\rm AJC}(t)} = \hat{H}_{\rm AJC}\ket{\phi_{\rm AJC}(t)}$. Using Eq.~\eqref{gintert}, we obtain the normalized solution of the AJC system as follows:
\begin{equation}\label{propto}
 \ket{\phi_{\rm AJC}(t)} = \frac{1}{\mathcal{N}_A}\pas{\hat{\mathcal{A}} \ket{\psi_{\rm JC}(t)}},
\end{equation}
where $\vert\mathcal{N}_A\vert^2 = |\mathcal{N}_0|^{-2}[\langle n_0 \rangle|\beta_{g}|^2+\pap{1+\langle n_0 \rangle}|\beta_{e}|^2]$ has been included because the SUSY intertwining operator (Eq.~\eqref{Loper}) is not unitary, 
$\langle n_0 \rangle$ being the initial mean photon number,  $\langle n_0 \rangle = \sum_{n=0}^{\infty}n\vert C_{n}\vert^2$. 
The process described above is schematically represented in Figure ~\ref{Diag_Map}.
It is noteworthy that to derive the solution $\ket{\phi_{\rm AJC}(t)}$ associated with $\hat{H}_{\rm AJC}$ one can either transform the (evolved) solution $\ket{\psi_{\rm JC}(t)}$, as indicated in Eq.~\eqref{propto}, or to start from the initial state $\ket{\psi_{\rm JC}(0)}$, then transform it into $\ket{\phi_{\rm AJC}(0)}$ using $\hat{\mathcal{A}}$, and subsequently evolve it [cf. Eq. (\ref{gintert})] via $\exp\pas{-it\hat{H}_{\rm AJC} }$. 
In turn, Eq.~\eqref{propto} allows for the direct computation of the expectation value of a given observable $\hat{\mathcal{O}}$ associated with the AJC system in the Schr\"odinger picture, as follows:
\begin{equation}\label{transfop}
\bra{\phi_{\rm AJC}(t)}\hat{\mathcal{O}}\ket{\phi_{\rm AJC}(t)}=
\frac{\bra{\psi_{\rm JC}(t)}\hat{\mathcal{A}}^{\dagger}\hat{\mathcal{O}}\hat{\mathcal{A}}\ket{\psi_{\rm JC}(t)}}{\vert\mathcal{N}_{A}\vert^2} .
\end{equation}
Thus the expectation values of observables associated with the AJC model can be obtained from the solution of the JC model by a remarkably easy transformation  
\begin{equation}\label{Eq_Oper_Trans}
\hat{\mathcal{A}}^{\dagger}\hat{\mathcal{O}}\hat{\mathcal{A}}=\begin{pmatrix}
\hat{a}\hat{\mathcal{O}}_{11}\hat{a}^\dagger & \hat{a}\hat{\mathcal{O}}_{12}\hat{a}\\
\hat{a}^{\dagger}\hat{\mathcal{O}}_{21}\hat{a}^{\dagger} & \hat{a}^{\dagger}\hat{\mathcal{O}}_{22}\hat{a}
\end{pmatrix},
\end{equation}
of the corresponding operator $\hat{\mathcal{O}}$, through the also simple intertwining operator~\eqref{Loper}. In this  representation, any arbitrary qubit operator $\hat{\mathcal{O}}_{\rm Qubit} \in \pac{\hat{\sigma}_\alpha, \hat{\sigma}_{\pm}}$, $\alpha \in \{x, y, z\}$, can be expressed in a simplified form as:
\begin{equation}\label{OQubit}
\hat{\mathcal{A}}^{\dagger}\hat{\mathcal{O}}_{\rm Qubit}\hat{\mathcal{A}}=\begin{pmatrix}
O_{11}\pap{\mathbb{I}+\hat{n}} & O_{12}\hat{a}^2\\
O_{21}\pap{\hat{a}^\dagger}^2 & O_{22}\hat{n} 
\end{pmatrix},
\end{equation}
where $O_{nm}\in \mathbb{C}$ are the matrix elements of $\hat{\mathcal{O}}_{\rm Qubit}$. 
In addition, for any field operator $\hat{\mathcal{O}}_{\rm Field}\in
\{ f_{\hat{n}},f_{\hat{a}}, f_{\hat{a}^{\dagger}} \}$, 
where $f_{\hat{z}}$ represents a function of the corresponding sub-index operator, Eq.~\eqref{Eq_Oper_Trans} can be written as:
\begin{equation}\label{Ofield}
\hat{\mathcal{A}}^{\dagger}\hat{\mathcal{O}}_{\rm Field}\hat{\mathcal{A}}=\begin{pmatrix}
\hat{a}\hat{\mathcal{O}}\hat{a}^{\dagger} & 0\\
0 & \hat{a}^{\dagger}\hat{\mathcal{O}}\hat{a}
\end{pmatrix}.
\end{equation}
Then, this mapping provides a general approach for obtaining expectation values of observables $\hat{\mathcal{O}}$ of the AJC model from the solution corresponding to the JC model employing a simple transformation on $\hat{\mathcal{O}}$. In the following subsections, we derive exact analytical expressions for calculating the expectation value of a general operator $\hat{\mathcal{O}}$ in each subsystem: the field and the qubit. 

\subsection{Computing expectation values of qubit operators}\label{Sec_Q_OPer_Cal}

In this section, we present a general framework for computing the expectation value of a conventional qubit operator by mapping the solutions of the JC to the AJC model.
To achieve this, we will take the general initial state given by Eq.~\eqref{gic} and let it evolve by the action of the evolution operator of the JC model  (Eq.~\eqref{U_I}). 
We first express a general qubit operator as $
\hat{\mathcal{O}}_{\rm Qubit}=\hat{\mathcal{O }}_{\rm Qubit}^{{\rm d}} +\hat{\mathcal{O}}_{\rm Qubit}^{{\rm ad}}$, where $\hat{\mathcal{ O}}_{\rm Qubit}^{{\rm d} }$ is a diagonal spin operator (e.g. $\hat{\sigma}_z$) and $\hat{\mathcal{O}}_{\rm Qubit}^{{\rm ad}}$ is an antidiagonal spin operator (e.g. $\hat{\sigma}_x$, $\hat{\sigma}_y$, $\hat{\sigma}_{+}$ or $\hat{\sigma}_-$). Then, using Eq. \eqref{OQubit}, we write the transformed operators as 
\begin{subequations}
\begin{align}
\hat{\mathcal{A}}^{\dagger}\hat{\sigma}_{z}\hat{\mathcal{A}}&=
\begin{pmatrix}
\mathbb{I}+\hat{n} & 0\\
0 & -\hat{n}
\end{pmatrix}
,\label{Dig_Qubit}\\
\hat{\mathcal{A}}^{\dagger}\hat{\mathcal{O}}_{\rm Qubit}^{ad}\hat{\mathcal{A}}&=
\begin{pmatrix}
0 & O_{12}\pap{\hat{a}}^2\\
O_{21}\pap{\hat{a}^{\dagger}}^2 & 0
\end{pmatrix}
.\label{ADiag_Qubit}
\end{align}
\end{subequations}
 In the last expressions, we used the fact that $\hat{\sigma}_z$ is the only diagonal qubit operator. By following the path of red arrows in Fig.~\ref{Diag_Map} and from the general initial state for the JC model given by Eq.~\eqref{gic}, we calculate the expectation value of $\hat{\sigma}_z$ for the AJC model as:
\begin{widetext}
\begin{equation}\label{Exp_5}
\bra{\phi_{\rm AJC}(t)}  \hat{\sigma}_z \ket{\phi_{\rm AJC}(t)} 
    =\frac{1}{|\mathcal{N}|^2}\sum_{n=0}^\infty |C_n|^2\pas{|\beta_{e}|^2 \mathcal{T}_{n+1}-|\beta_{g}|^2 \mathcal{T}_{n}}
    + 4\sqrt{\pap{n+1}^{3}}\ \mathrm{Re}\pas{\beta_{g} \beta_{e}^*C_{n+1}C_n^*\mathcal{G}_{n+1}\mathcal{F}_{n+1}},
\end{equation}
\end{widetext} 
where $|\mathcal{N}|^2=\langle n_0 \rangle|\beta_{g}|^2+\pap{1+\langle n_0 \rangle}|\beta_{e}|^2$. The transition amplitudes 
\begin{equation}\label{tranampl}
\mathcal{T}_{n}=n\pas{|\mathcal{F}_{n}|^2-n |\mathcal{G}_{n}|^2}
\end{equation} 
also appear in Eq.~\eqref{Exp_5}, an equation that can be drastically simplified if the initial state of the matter subsystem is, for instance, the ground state $|\beta_g|^2 = 1$, $\beta_e = 0$. In this case, the last two terms of Eq.~\eqref{Exp_5} vanish, and the expectation value takes on a reduced expression. Following a similar approach as described above, but using Eq.~\eqref{ADiag_Qubit}, we now compute the general explicit form of the expectation value of any antidiagonal qubit operator, $\hat{\sigma}_x$ or $\hat{\sigma}_y$, which can be written as linear combinations of $\hat{\sigma}_{+}$ and $\hat{\sigma}_{-}$. Without loss of generality, we focus our attention on the operator $\hat{\sigma}_{+}$, obtaining by direct calculation
\begin{widetext}
\begin{equation}\label{Sigma_mas_SUSY}
\begin{split}
\bra{\phi_{\rm AJC}(t)}  \hat{\sigma}_{+} \ket{\phi_{\rm AJC}(t)} 
=&\frac{1}{|\mathcal{N}|^2}\sum_{n=0}^\infty\bigg(\sqrt{n+1}C^*_{n}C_{n+1}\mathcal{F}_{n+1}\bigg[(n+2)\mathcal{G}_{n+2}|\beta_{e}|^2-n\mathcal{G}_{n}|\beta_{g}|^2\bigg]\\
    &+  \sqrt{(n+1)(n+2)}C^*_{n}C_{n+2}\mathcal{F}_{n+1}\mathcal{F}_{n+2}\:\beta_{g}\beta_{e}^{*}-  n(n+1)|C_n|^2\mathcal{G}_{n}\mathcal{G}_{n+1}
\beta_{g}^*\beta_{e}\bigg).
\end{split}
\end{equation}
\end{widetext}
From here, we can easily determine the expectation value of $\hat{\sigma}_{-}$, by evaluating the complex conjugate of Eq.~\eqref{Sigma_mas_SUSY}, considering that $\mathcal{G}_{n}^{*}=-\mathcal{G}_{n}$.  Similar to the expectation value of the qubit population (Eq.~\eqref{Exp_5}), the expectation value of the ladder qubit operator takes a simple form when the qubit is initialized in a simpler state. We will perform a similar analysis for the field operators in the next subsection, focusing on computing general expressions for the high-order expectation values of the field operators.

\subsection{Analysis of expectation values for field operators}\label{sec_field_op}

In quantum optics, the ability to calculate powers of the photon number operator $\langle \hat{n}^{k}\rangle$, creation operator $\langle (\hat{a}^\dagger)^{k}\rangle$, and annihilation operator $\langle \hat{a}^{k}\rangle$ is highly valued as through them a deeper understanding of the quantum properties of light and its interactions with matter is achieved. Moreover, the photon statistical properties are analyzed within the framework of the full counting statistics, providing insights into the photon number distribution and correlations.  These results can be directly measured using traditional quantum optics setups, facilitating experimental verification of theoretical predictions~\cite{Agarwalla_PRB19, Nesterov_PRA20, Zenelaj_PRB22, Xu_PRB13}. 
Below,  we show how the SUSY map allows the calculations of powers for these operators.

\subsubsection{Full counting statistics of photons}

The statistical properties of photons are analyzed within the framework of full counting statistics, which provides information about the probability distribution $P(n, t)$ of having $n$ photons within a time $t$. Instead of directly computing the probability distribution $P(n, t)$, it is more convenient to examine the behavior of the cumulants $\langle \hat{n}^k \rangle$, $k = 1,2\dots$ of the distribution. These cumulants offer valuable information about the statistical properties of photons, which are essential for understanding various quantum phenomena and their experimental manifestations. The $k$-th cumulant of the distribution can be expressed using SUSY transformations as follows: 
\begin{equation}
\hat{\mathcal{A}}^{\dagger}\hat{n}^k\hat{\mathcal{A}}= 
\begin{pmatrix}
(\mathbb{I}+\hat{n})^k & 0\\
0 & \hat{n}(\hat{n}-\mathbb{I})^k
\end{pmatrix}.
\end{equation}
Therefore, employing an approach similar to that described in Sec.~\ref{Sec_Q_OPer_Cal}, we explicitly calculate the form of the $k$-th cumulant of the photon distribution. After a simple calculation, we finally obtain
\begin{widetext}
\begin{equation}\label{Exp_nk}
   \begin{split}
\bra{\phi_{\rm AJC}(t)}  &     \hat{n}^{k} \ket{\phi_{\rm AJC}(t)} 
       =\frac{1}{|\mathcal{N}|^2}
       \sum_{n=0}^\infty |C_n|^2
       \bigg\lbrace|\beta_{e}|^2
       \pas{\pap{1+n}^{k+1}|\mathcal{F}_{n+1}|^2+(1+n)^2 n^k|\mathcal{G}_{n+1}|^2}
       \\
       &+|\beta_{g}|^2\pas{n\pap{n-1}^{k}|\mathcal{F}_{n}|^2+  n^{k+2}|\mathcal{G}_{n}|^2}\bigg\rbrace-2\sqrt{\pap{n+1}^3}
       \pas{n^k-\pap{1+n}^k}
       \mathrm{Re}\pas{\beta_{g} \beta_{e}^*C_{n+1}C_n^*\mathcal{G}_{n+1}\mathcal{F}_{n+1}},
   \end{split}
\end{equation}
\end{widetext}
a result that offers a powerful method for computing high-order cumulants for the AJC model over time.  For example, it provides an easy way to evaluate the Fano factor: ${\rm FF}={\rm Var}(\hat{n})/\langle \hat{n}\rangle$, where ${\rm Var}(\hat{n})=\langle \hat{n}^2\rangle-\langle \hat{n}\rangle^2$. Recall that ${\rm FF}=1$ means a coherent state, while values less (greater) than 1 indicate a sub-Poissonian (super-Poissonian) photon state. For completeness, in the following, we extend our analysis by calculating the SUSY map for the high-order cumulants of the creation and annihilation photon operators. This calculation will comprehensively understand the photon statistics and further clarify the relationship between the JC and the AJC models.

\subsubsection{Comprehensive analysis of field operator full counting statistics}

The statistical characteristics of the resulting photon field amplitude $\langle \hat{a}\rangle$ 
can be measured in contemporary experimental setups of light-matter interaction systems. 
The measuring method depends on the specific cavity quantum electrodynamics setup used. 
Potential approaches include Ramsey interferometry (where a second atom is employed to probe the cavity's state), homodyne measurement of photons escaping from the cavity by interfering them with a reference beam, and homodyne measurement conducted within the cavity by detecting the interfering fields with a second atom~\cite{Haroche_RMPhys01, Haroche_PRL03}. We begin by writing the general form of the SUSY map for the field amplitudes as: 
\begin{subequations}
\begin{align}\label{SUSY_ak}
\hat{\mathcal{A}}^{\dagger}\hat{a}^k\hat{\mathcal{A}}&= 
\begin{pmatrix}
\hat{a}^k\pap{\mathbb{I}+\hat{n}} & 0\\
0 & \hat{n}\hat{a}^k
\end{pmatrix},
\\
\hat{\mathcal{A}}^{\dagger}\pap{\hat{a}^\dagger}^k\hat{\mathcal{A}}&= 
\begin{pmatrix}
\pap{\mathbb{I}+\hat{n}}\pap{\hat{a}^\dagger}^k & 0\\
0 & \pap{\hat{a}^{\dagger}}^k\hat{n}
\end{pmatrix}
.\label{SUSY_adk}
\end{align}
\end{subequations}
These are related to the $k$-th cumulant of the field amplitude, which is directly associated with fluctuations in photon creation and annihilation. Similar to the qubit case, we define new transition amplitudes as follows:
\begin{equation}
\begin{split}
\mathcal{T}^{\tilde{m}}_{m}&=m\pas{\mathcal{F}_{\tilde{m}}\mathcal{F}_{m}^{*}+\tilde{m}\mathcal{G}_{\tilde{m}}\mathcal{G}_{m}^{*}},\\
\tilde{T}_{m}^{\tilde{m}}&=\tilde{m}\mathcal{F}_{m}\mathcal{G}_{\tilde{m}}-m\mathcal{F}_{\tilde{m}}\mathcal{G}_{m}, \label{tranampltilde}\\
\bar{\mathcal{T}}_{m}^{\tilde{m}}&=\mathcal{F}_{m}^{*}\mathcal{G}_{\tilde{m}}-\mathcal{F}_{\tilde{m}}^{*}\mathcal{G}_{m}.
\end{split}
\end{equation}
Given the general form of the initial state (see Eq.~\eqref{gic}) and the structure of the SUSY transformation Eq.~\eqref{SUSY_ak}, and using the following relation: $\hat{a}^k\ket{n}=\sqrt{n!/\pap{n-k}!}\ket{n-k}$, we obtain the $k$-th cumulant:
\begin{widetext}
\begin{equation}\label{Eq_ak_2}
   \begin{split}
\bra{\phi_{\rm AJC}(t)}  \hat a^k \ket{\phi_{\rm AJC}(t)} 
       =\frac{1}{|\mathcal{N}|^2}&\sum_{n=0}^\infty \sqrt{\frac{(n+k)!}{n!}} C_n^* \bigg(C_{n+k}\bigg[|\beta_{e}|^2\mathcal{T}_{n+k+1}^{n+1}+|\beta_{g}|^2 \mathcal{T}_{n}^{n+k}\bigg]\\
       &+\pas{\beta_{g} \beta_{e}^* \:\sqrt{n+k+1}C_{n+k+1} \tilde{\mathcal{T}}_{n+1}^{n+k+1} +\beta_{g}^* \beta_{e}\:n\sqrt{n+k}C_{n+k-1}\bar{\mathcal{T}}_{n}^{n+k}}\bigg).
   \end{split}
\end{equation}
\end{widetext}
Furthermore, as in the case of the qubit, the expectation value of $(a^\dagger)^k$ can be obtained simply from a complex conjugation of Eq.~\eqref{Eq_ak_2}. In the next section, some specific numerical examples are given to show the usefulness of the results that have been found.

\section{Numerical experiments}\label{Sec_Num_Exp}

In this section, we will present examples that clarify the applicability of the general results developed in Section~\ref{Sec_Exact_Dyn}, analyzing the evolution in time of some observables of the AJC system. 
The initial state of the partner JC system is taken to be $\ket{\psi_{\rm JC}(0)}=\ket{{\rm Qubit}}\otimes\ket{{\rm Field}}$, with
\begin{equation}\label{qubit_state}
\ket{{\rm Qubit}}=\cos\theta\ket{g}+e^{i \varphi}\sin\theta\ket{e},
\end{equation}
where the initial probability amplitudes are given by $|\beta_e|^2=\sin^2\theta$ and $|\beta_g|^2=\cos^2\theta$.  The angle $\theta$ modulates the ratio of the probability amplitudes of the atom to be initially in the ground or the excited state, and the angle $\varphi$ shifts the phase between both states of the qubit (from now on we set $\varphi=\pi/4$).
On the other hand, we initialize the field subsystem so that 
\begin{equation}\label{photon_state}
\ket{{\rm Field}}=\frac{1}{N_{\vartheta}}\pas{\ket{\alpha}+e^{i\vartheta}\ket{-\alpha}},
\end{equation}
is a Schr\"odinger cat-like state, i.e. a superposition of the two coherent states $\ket{\alpha}$ and $\ket{-\alpha}$, the normalization constant being 
$N_{\vartheta}^2 = 2\left(1+ e^{-2|\alpha|^2}\cos\vartheta \right)$. For simplicity, from now on we assume that $\alpha$ is real.
Field states as Eq.~\eqref{photon_state} allows to generate states like the following: (a) {\it Even coherent cat state} ($\vartheta=0$): the photon number distribution is nonzero only for even photon numbers and the average photon number is $\langle n_0\rangle=|\alpha|^2\tanh\left(|\alpha|^2\right)$.  (b) {\it Odd coherent cat state} ($\vartheta=\pi$): only odd number of photons have a nonzero probability, and the average photon number is $\langle n_0\rangle=|\alpha|^2\coth\left(|\alpha|^2\right)$. (c) {\it Yurke-Stoler coherent state} ($\vartheta=\pi/2$): the average photon number is $|\alpha|^{2}$.  In addition, field states such as those in Eq.~\eqref{photon_state} are accessible light states that can currently be created in the laboratory. From now on we set $\vartheta=0$, so the initial state of the photon in Eq.~\eqref{photon_state} becomes an even cat state, whose coefficients in the basis of the Fock states $\{ |n\rangle\}$ are given by
\begin{equation}
C_{n}=\frac{1}{N_{0}}\exp\pas{-\frac{|\alpha|^2}{2}}\frac{\pap{1+\pap{-1}^n}}{\sqrt{n!}}\ \alpha^n.
\end{equation}
Next, and for the states that we have just discussed, we show various numerical results that have been computed from the analytical expressions obtained in Section~\ref{Sec_Exact_Dyn}. In turn, these analytical results were tested numerically using the Quantum Toolbox in Python (QuTiP)~\cite{Johansson2013}.  We have set the time scale in units of $\omega_c$ and have considered a weak coupling regime, that is, we set $\lambda = 0.1\,\omega_c$. Furthermore, we have taken $\alpha = 4$.  A convergence analysis has been carried out for both the numerical and analytical calculations. The Fock basis was truncated for $N=250$ photons in both cases. \\

It is worth mentioning that, for numerical calculations, the path of the blue arrows shown in Figure~\ref{Diag_Map} has been followed; note that the initial condition of the AJC system is not, in general, the same as the initial condition for the JC model, but rather its image under the mapping defined by $\hat{\mathcal{A}}$. Thus, we have started from the initial condition of the JC model and, after transforming it with the operator in Eq.~\eqref{Loper}, its evolution was obtained by $\hat U_{AJC}(t)$. This, on the one hand, validates the commutative diagram of Figure~\ref{Diag_Map}, remembering that the analytical results were obtained taking the path of the red arrows: we have used the time-dependent solution for the JC system. On the other hand, the numerical results allow us to verify the accuracy and robustness of the obtained analytical results. Furthermore, it is also worth noting that the analytical results presented are valid out of resonance, that is $\Delta\neq 0$. For the resonant case $\Delta=0$ of a coherent state and an atom initially in the ground state, the results are shown in Appendix~\ref{apnedix_1}. Next, we present the analysis of the qubit subsystem in Sec.~\ref{Results_Qubit_Nume} and of the light subsystem in Sec.~\ref{Results_Boson_Nume}.

\begin{figure*}[t!]
\centering
\includegraphics[width=0.9\linewidth]{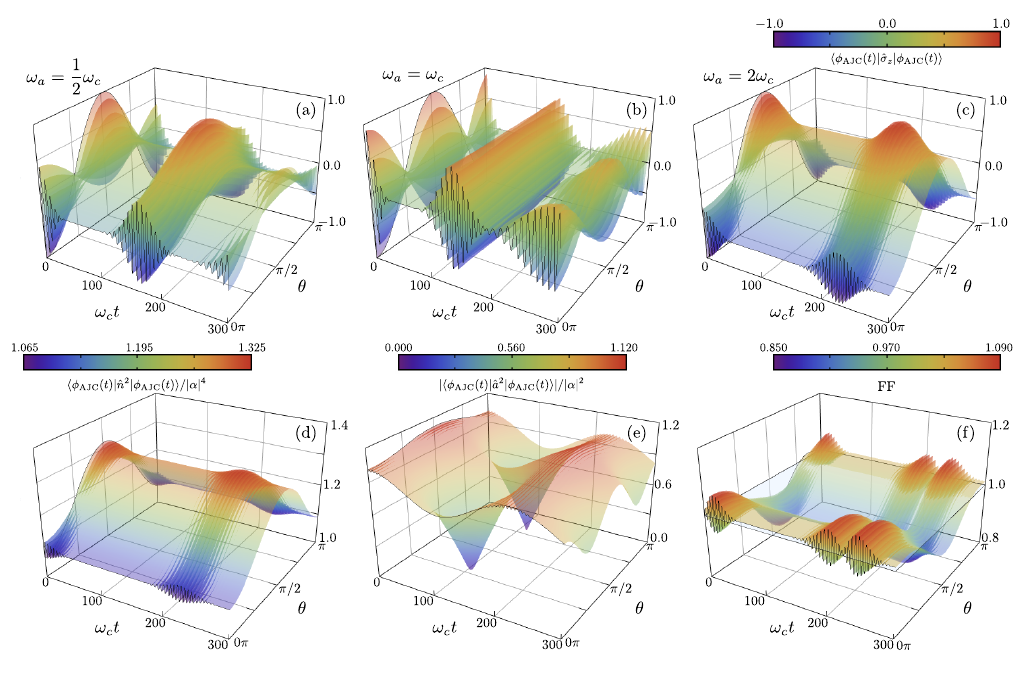}
\caption{\label{fig_3}  {\bf Exact dynamic landscapes of the AJC model via SUSY transformation.}  In the upper panels, we present the results for the qubit subsystem. 
The dynamics of the qubit, given by the expected value of $\hat{\sigma}_z$, is represented as a function of the angle of the initial state of the qubit $\theta$ and the evolution time, for three values of the qubit frequency: (a) $\omega_a < \omega_c$, (b) $\omega_a = \omega_c$ and (c) $\omega_a > \omega_c$.
In the bottom panels, we show the fingerprints for the full photon counting statistics.
Panels (d) and (e) show the evolution of fluctuations in photon number and field amplitude, respectively.
In panel (f) we illustrate the complex trade-off from sub-Poissonian to super-Poissonian statistics (see blue plane for ${\rm FF}=1$), presented by the Fano factor.
The results for the field subsystem (bottom panels) correspond to the case of $\omega_a > \omega_c$, presenting a qualitatively similar behavior in the other regimes though.}
\end{figure*}

\subsection{Atomic inversion}\label{Results_Qubit_Nume}

The first observable we are going to analyze is the atomic inversion (see Eq.~\eqref{Exp_5}), defined as the difference between the probabilities of finding the two-level system in the excited or the ground state. Thus, the atomic inversion gives information about the state of the atomic subsystem.  Note that the expression of Eq.~\eqref{Exp_5} reduces in the current situation, as $C_n^*C_{n+1} = 0$ for the initial even (odd) cat state. The first row of Figure~\ref{fig_3} shows the atomic inversion of the AJC system for three specific cases of the atomic frequency  $\omega_a$: (a) $\omega_a<\omega_c$,  (b) $\omega_a=\omega_c$ and (c) $\omega_a>\omega_c$.  The archetypal behavior of collapses and revivals can be appreciated, particularly at $\theta = 0$. Observe, however, the disappearance for $\theta =\pi/4$ ($\theta =3\pi/4$) of the oscillations before ($\omega_ct\sim 0$) and after ($\omega_ct\sim 300$) the first revival. This can be particularly seen in Fig~\ref{fig_3}(a).\\

Another interesting fact is that at resonance (Figure~\ref{fig_3}(b)) the first revival retains its amplitude, regardless of the initial atomic state (without considering $\theta$).  As in the previous case, the oscillations before and after the first revival (i.e., the second revival) can be manipulated by judiciously controlling the initial state of the two-level system ($\theta=\pi/4$, for example), which is possible with the currently available laboratory technology. 
Note also that, in resonance, the first and second collapses approach each other. On the other hand, out of resonance (Figures~\ref{fig_3}(a) and~\ref{fig_3}(c)) atomic transitions are not very probable despite the collapse-revival behavior. Thus, it is possible to perform collapse-revival engineering by controlling the initial atomic state, as well as the system detuning, in the AJC system.

\subsection{Photon full-counting statistics}\label{Results_Boson_Nume}
The analysis of the higher-order cumulants of the field amplitude distribution described in Section~\ref{sec_field_op} is now presented. A significant understanding of the characteristics of the field subsystem is obtained, since the quantities presented are measurable and contain information about the quantum character of the electromagnetic field.\\

The square of the photon number operator provides information about the state of the field, specifically about the mean number of photons present in it. 
Usually, the analysis of the mean photon number is obtained directly from $\langle\hat n\rangle$, however, the square of the number operator carries equivalent information. 
Furthermore, $\langle\hat n^2\rangle$ gives information about the quantum (classical) nature of the field, through the Fano factor, as explained before. Figure~\ref{fig_3}(d) shows the expectation value of the square of the number operator, according to Eq.~\eqref{Exp_nk}, for $k=2$ and $\omega_a = 2\omega_c$. As in the case of the atomic inversion, the Eq.~\eqref{Exp_nk} is simplified for the chosen initial state of the radiation: a Schr\"odinger's cat type. Note that $\langle\hat n^2\rangle$ is scaled by a factor of $1/|\alpha|^4$ through the color bar in Figure~\ref{fig_3}(d).  Collapses and revivals can be observed, along with the disappearance of the oscillations at the beginning of the interaction (before the first revival), as well as after the first revival (not shown). 
However, the mean value of the square of the photon number takes larger values for the AJC system, compared to the JC model (not shown either), due to the counter-rotating terms $\pac{\hat a^\dagger \hat\sigma_+,\hat a \hat\sigma_-}$ present in $\hat H_{AJC}$.\\

Analogously, the expectation value of the square of the annihilation operator is analyzed. It is directly connected to the field quadratures and can therefore be measured in the laboratory. 
Figure~\ref{fig_3}(e) shows the evolution of the expectation value of the square of the annihilation operator, according to Eq.~\eqref{Eq_ak_2}. Note that for the initial even (odd) cat state and $k$ an even number, the last two terms in Eq.~\eqref{Eq_ak_2} disappear. 
In addition, as $k = 2$, we have that 
\begin{equation}
C_n^*C_{n+2} = \alpha^2\, e^{-|\alpha|^2}\, \frac{|\alpha|^{2n}}{\sqrt{n!(n+2)!}}[1+(-1)^n]^2.
\end{equation}
The same parameters as for the square of the photon number operator (Figure~\ref{fig_3}(d)) are considered in Figure~\ref{fig_3}(e), and $\langle\hat a^2\rangle$ is scaled by a factor of $1/|\alpha|^2$. 
The oscillations shown in Figure~\ref{fig_3}(e) are a different manifestation of the interference between the superposition of Fock states that generate the collapses and revivals in $\langle\hat\sigma_z\rangle$ and $\langle\hat n^2\rangle$, which are directly detectable by homodyne measurements.  Indeed, such oscillations are a fingerprint of entanglement between the atomic and radiation subsystems, just like collapse-revival behavior, and decrease in amplitude over time. In addition, such amplitude can be precisely adjusted only by properly establishing the initial atomic state through the control parameter $\theta$, as can be seen in Figure~\ref{fig_3}(e).\\

The FF is another quantity of interest that can be straightforwardly obtained from our results, more precisely from Eq.~\eqref{Exp_nk}.  As mentioned, FF provides us with information about the nature of the field.   In particular, ${\rm FF}<1$ means that the field is sub-Poissionian and presents quantum features; ${\rm FF} = 1$ means that the field is Poissonian, that is, that the field is in a coherent (quasi-classical) state; and
${\rm FF}>1$ means that the field presents a super-Poissonian distribution of photons. The temporal evolution of the FF is shown in Fig.~\ref{fig_3}(f) for the same parameter values used in Fig.~\ref{fig_3}(d)(e).  A horizontal plane at ${\rm FF} = 1$ has been added, as a reference for the eye.  It can be seen that the field photon distribution oscillates between Poissonian, sub-Poissonian, and super-Poissonian over time.  However, such time intervals can be engineered by controlling the initial atomic state through $\theta$.  In particular, for $\theta = \pi/2$, namely, an initial two-level system in the excited state, ${\rm FF}<1$ for all $t$, which reveals the quantum nature (sub-Poissonian) of the electromagnetic field.\\

Therefore, it is also possible to control the quantum features of the field in the AJC system by appropriately fixing the initial atomic state. It is worth mentioning that the simpler case of a coherent state is usually shown for the AJC system. However, we have shown that the more involved case of an even cat state for the field shows similar features, such as well-defined collapses and revivals in atomic inversion and mean photon number.

\section{Conclusions}
In this study, we have presented a new interpretation of a model widely studied in quantum optics, the Jaynes-Cummings model.  Such interpretation lays on the discovery that a SUSY transformation can map the JC model into the anti-Jaynes-Cummings model.  Moreover, the exact dynamics can be mapped through this SUSY transformation, facilitating a simplified description to understand the dynamics arising from a very general interaction state between light and matter. We have provided a comprehensive analysis, both theoretical and numerical, of the dynamics of the AJC model as a SUSY partner of the JC model. We extended these results to the context of full counting statistics of photons.  Our findings allow the computation of exact solutions regardless of the initial state. However, an appropriate choice of the initial conditions of the system makes it possible to engineer the collapse-revival behavior and the quantum characteristics of the interacting radiation field. We have evaluated and analyzed the dynamics of the Fano factor as a fingerprint of complex quantum behavior over time.

The results presented in this study can be experimentally validated using current quantum technologies and are of interest in hot areas of physics such as quantum optics, quantum communication, information processing, and quantum computing.  Looking ahead, this work opens new avenues to explore connections between more elaborated models in quantum optics, such as the Dicke model, and their corresponding SUSY partners.

\section*{Acknowledgments}
The authors thank Diego Bussandri for careful reading, as well as valuable comments, on the original manuscript. This work was supported by Spanish MCIN with funding from European Union Next Generation EU (PRTRC17.I1) and Consejeria de Educacion from Junta de Castilla y Leon through QCAYLE project, as well as grants PID2020-113406GB-I00 MTM funded by AEI/10.13039/501100011033, and RED2022-134301-T.  I.A.B-G acknowledge CONAHCyT (M\'exico) for financial support through project A1-S-24569, and IPN (M\'exico) for supplementary economical support through project SIP20232237.

\appendix
\section{Further factorization properties for the intertwining operators}\label{apnedix_01}
In addition to the analytical techniques discussed throughout the manuscript, we would like to mention some straightforward general consequences of the intertwining relation given by Eq.~\eqref{L_oper_JC}. First, the following intertwining relation is also satisfied
\begin{equation}
     \hat{H}_{\rm JC}\pap{\omega_a}\hat{\mathcal{A}}^{\dagger} =\hat{\mathcal{A}}^\dagger \hat{H}_{\rm AJC}\pap{\omega_a-2\omega_c}.
\end{equation}
Therefore, it can be checked that $\hat{\mathcal{A}}^\dagger\hat{\mathcal{A}}$ is a Hermitian symmetry of $\hat{H}_{\rm JC}$. We can express this symmetry as $[\hat{\mathcal{A}}^\dagger\hat{\mathcal{A}},\hat{H}_{\rm JC}]= 0$. In general, we define an eigenvalue equation $\hat{\mathcal{A}}^\dagger\hat{\mathcal{A}}|\Psi_n\rangle = \lambda_n|\Psi_n\rangle$, where the eigenvalues $\lambda_n = n$, with $n=0,1,\dots$, and the corresponding eigenvectors $|\Psi_n\rangle$ are given by
\begin{equation}
    |\Psi_n\rangle= \begin{cases} 
    |g,0\rangle,& \text{for } n=0,\\
    \left\{|e,n-1\rangle  , |g,n\rangle  \right\},& \text{for } n\geq 1.
    \end{cases}
\end{equation}
Similarly, the Hermitian symmetry for the AJC model can be expressed as $[\hat{\mathcal{A}}\hat{\mathcal{A}}^\dagger,\hat{H}_{\rm AJC}]= 0$, and the corresponding eigenvalue equation is $\hat{\mathcal{A}}\hat{\mathcal{A}}^\dagger|\Phi_n\rangle = \lambda_n|\Phi_n\rangle$, where $\lambda_n$ is the same as in the JC model symmetry. The corresponding eigenvectors are given by
\begin{equation}
    |\Phi_n\rangle= \begin{cases}
    |e,0\rangle, & \text{for }n = 0,\\
    \left\{|e,n\rangle,|g,n-1\rangle\right\},&\text{for }  n\geq 1.
    \end{cases}
\end{equation}
Note that in both cases the ground state is a singlet, while the excited states are two-fold degenerate. Furthermore, the intertwining operators $\mathcal{\hat A}$ and $\mathcal{\hat A}^\dagger$ annihilate the respective ground states $|\Psi_0\rangle$ and $|\Phi_0\rangle$ which in turn allows the construction of coherent states for the AJC (JC) system, nevertheless this is out of the scope of the present work. The  Hamiltonians can be diagonalized within each of the eigenspaces above described. In addition, for any $|\psi_n\rangle = c_1|\Psi_n\rangle + c_2|\Psi_n^\prime\rangle$ and $|\phi_n\rangle = d_1|\Phi_n\rangle + d_2|\Phi_n^\prime\rangle$, $c_i,d_i\in\mathbb C$, $i = 1,2$, where $|\Psi_n^\prime\rangle$, $|\Phi_n^\prime\rangle$ denote solutions (if any) that are linearly-independent of $|\Psi_n\rangle$, $|\Phi_n\rangle$, respectively, it follows that: $\mathcal{\hat A}|\psi_n\rangle = \sqrt{n}|\phi_n\rangle$.

\section{Analytical reductions: transitioning from resonant JC model to non-resonant AJC model via SUSY}\label{apnedix_1}
\begin{figure*}[t!]
\centering
\includegraphics[width=0.9\linewidth]{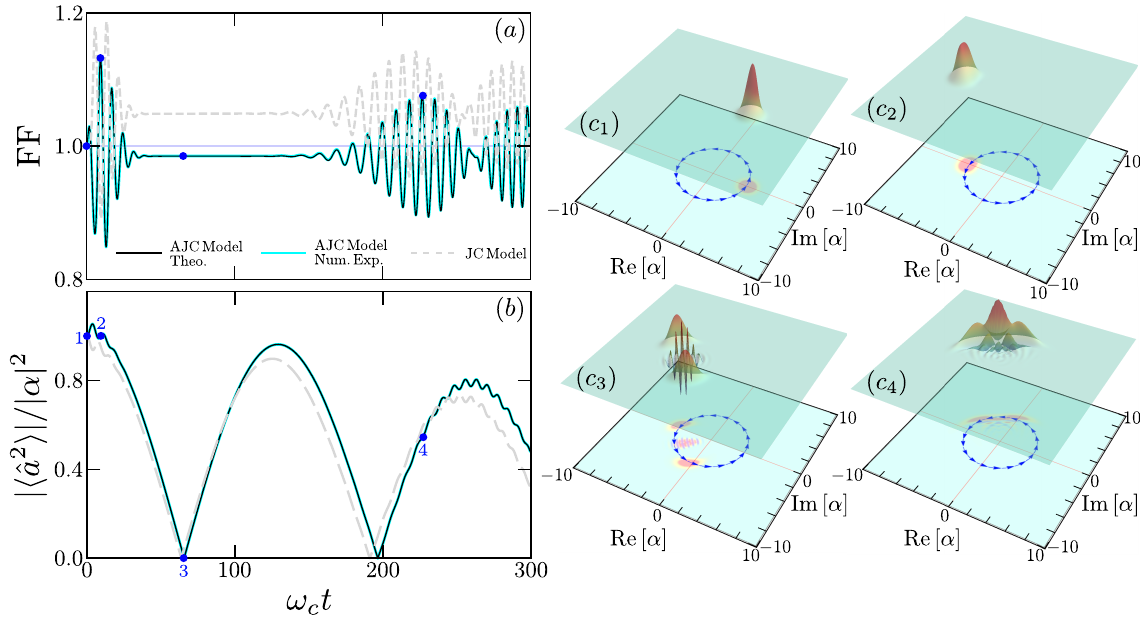}
\caption{\label{fig_4}  {\bf 
Parallelism of higher-order cumulants for the photon probability distribution for the JC and AJC models.} In the resonance limit of the JC model and with an initial state given by $\ket{\psi_{\text{JC}}(0)}=\ket{g}\otimes \ket{\alpha}$, being $\ket{\alpha }$ a coherent state with $\alpha=4$, we characterize the behavior of the higher-order cumulants of the time-dependent photon number distribution using (a) the Fano factor and (b) the fluctuation of the amplitude of the field.  In panel (a), a horizontal blue line at ${\rm FF}=1$ is plotted to delineate the transition from sub-Poissonian to super-Poissonian photon statistics.
Additionally, we denote four instants of time with blue dot markers labeled 1 to 4 in both the Fano factor and the field amplitude fluctuation.
In panel (c), the Wigner function $W(\alpha,\alpha^*)$ is plotted for the four selected times.
The density plot corresponds to the projection of the Wigner function on the plane $(\alpha,\alpha^*)$, with a blue circle of radius $4$, and arrows indicating the direction of rotation and evolution of the Wigner function.}    
\end{figure*}
The general approach described in the main text allows us to characterize the dynamics of the AJC model starting from the solution of the JC model. Recall that the mapping defined by the SUSY transformation produces an AJC model with a qubit frequency shifted by the field frequency. Here, we will delve into the scenario where the Hamiltonian of the JC model is in resonance, i.e., $\omega_c=\omega_a=\omega$. 
Consequently, the corresponding SUSY partner generated from Eq.~\eqref{h_SUSY_AJC} is given by:
\begin{equation}
\hat{H}_{\rm AJC} = \omega \hat{a}^{\dagger}\hat{a}-\frac{\omega}{2}\hat{\sigma}_z +\lambda\pap{\hat{a}^{\dagger}\hat{\sigma}_{+}+ \hat{a}\hat{\sigma}_{-}}.
\end{equation}
In the current resonance limit ($\Delta=0$), all the equations presented in the main text can be simplified. 
To illustrate this, we will consider a particularly simple case where the initial state is $\ket{\psi_{\rm JC}(0)}=\ket{g}\otimes \ket{\zeta}$, with $\ket{\zeta}=\sum_{n=0 }^{\infty}C_{n}\ket{n}$.
Observe that the normalization constant becomes $|\mathcal{N}|^2=\langle n_0 \rangle$. Therefore, the complete dynamics of the expectations values for both subsystems is given by: 

\begin{widetext}
\begin{subequations}\label{Eq_g_Reso}
\begin{align}
\bra{\phi_{\rm AJC}(t)}  \hat{\sigma}_{+} \ket{\phi_{\rm AJC}(t)} 
&=\frac{i}{\langle n_0 \rangle}\sum_{n=0}^\infty \sqrt{n\pap{n+1}}\cos\pas{\lambda\sqrt{n+1}t} \sin\pas{\lambda\sqrt{n}t}\;C^*_{n}C_{n+1}.\\
\bra{\phi_{\rm AJC}(t)}  \hat{\sigma}_z  \ket{\phi_{\rm AJC}(t)} 
    &=-\frac{1}{\langle n_0 \rangle}\sum_{n=0}^\infty n \cos\pas{2\lambda\sqrt{n}\;t}\;|C_n|^2.\\
\bra{\phi_{\rm AJC}(t)}  \hat{n}^{k}  \ket{\phi_{\rm AJC}(t)} 
&=\frac{1}{\langle n_0 \rangle}\sum_{n=0}^\infty |C_n|^2\pas{n\pap{n-1}^{k}\cos^2\pas{\lambda\sqrt{n}t}+  n^{k+1}\sin^2\pas{\lambda\sqrt{n}t}}.\\
\begin{split}
\bra{\phi_{\rm AJC}(t)}  \hat a^k \ket{\phi_{\rm AJC}(t)} 
&=\frac{1}{\langle n_0 \rangle}\sum_{n=0}^\infty \sqrt{\frac{(n+k)!}{n!}} C_n^* C_{n+k} \bigg(
n\cos\pas{\lambda\sqrt{n+k}\;t}\cos\pas{\lambda\sqrt{n}\;t}\\
&\quad +\sqrt{n\pap{n+k}}\sin\pas{\lambda\sqrt{n+k}\;t} \sin\pas{\lambda\sqrt{n}\;t}\bigg).
\end{split}
\end{align}
\end{subequations}
\end{widetext}

To obtain these expressions we have used the  transition amplitudes Eq.~\eqref{tranampl} and Eq.~\eqref{tranampltilde}, given in the present case by 
\begin{align}
\mathcal{T}_{n}  &= n \cos\pas{2\lambda\sqrt{n}\;t}, \\
\mathcal{T}^{\tilde{m}}_{m} &= m\cos\pas{\lambda\sqrt{\tilde{m}}\;t}\cos\pas{\lambda\sqrt{m}\;t}\nonumber \\
&\quad +\sqrt{m\tilde{m}}\sin\pas{\lambda\sqrt{\tilde{m}}\;t} \sin\pas{\lambda\sqrt{m}\;t}. 
\end{align}
Note that expressions Eq.~\eqref{Eq_g_Reso} depend only on the coefficients $C_n$ of the initial state of the field. Here we simplify the notation as $\bra{\phi_{\rm AJC}(t)} \hat{O} \ket{\phi_{\rm AJC}(t)}=\langle \hat{O}\rangle_{\phi_t}$. 
For the Fock state $\ket{m}$, the coefficients satisfy $C_{n}=\delta_{nm}$ and $\langle n_0 \rangle = m$.  Therefore, the expectation values are easily seen to be $\langle \hat{\sigma}_{+} \rangle_{\phi_{t}} =\langle\hat a^k\rangle_{\phi_{t}}=0$,  and $\langle \hat{\sigma}_z \rangle_{\phi_{t}}=-\cos \pas{2g\sqrt{m}\;t}$. Finally, the mean value of the $k$-th cumulant for the number of photons is shown to be $\langle \hat{n}^{k} \rangle_{\phi_{t}} =\pap{m-1}^{k}\cos^2\pas{g\sqrt{m}t}+  m^{k}\sin^2\pas{g\sqrt{m}t}.$ The FF for a Fock state $|m\rangle$ can be straighforwardly calculated:
\begin{equation}\label{FF_Fock}
{\rm FF}_{m}\pap{t}=\frac{\sin^2\pas{2g\sqrt{m}t}}{2\pap{2m-\cos\pas{2g\sqrt{m}t}-1}}.
\end{equation}
This function exhibits a distinctive behavior: it always remains less than one. 
Specifically, for $m = 1$, the Fano factor decreases as ${\rm FF}_{1} = \cos^2(gt)$. 
When $m > 1$, the Fano factor can be approximated as ${\rm FF}_{m} \sim \sin^2\pas{2g\sqrt{m}\left(t+\frac{1}{4}\right)}/2(2m-1)$. Therefore, the distribution of photons presents a sub-Poissonian (purely quantum) behavior.\\

On the other hand, if the initial state of the field is a coherent state $\ket{\alpha}$, the coefficients $C_n$ are given by $C_n = e^{-|\alpha|^2/2}\frac{\alpha^n}{\sqrt{n!}}$. We also have $C_{n}^{*}C_{n+k}=\alpha^k \exp\pas{-|\alpha|^2}\frac{|\alpha|^{2n}}{\sqrt{n!\pap{n+k}!}}$, and $\langle n_0 \rangle = |\alpha|^2$.  Exact expressions for the expectation values can be obtained using Eq.~\eqref{Eq_g_Reso}.  In Figure~\ref{fig_4}, we characterize the high-order cumulants of the photon distribution for both the JC model and its SUSY partner, the AJC model, for $|\psi_{\rm JC}(0)\rangle = |g\rangle\otimes|\alpha\rangle$.  Due to the simple choice of the initial state, the action of the SUSY transformation leaves the state invariant.  For this reason, a direct comparison between both models is possible.\\

In Fig.~\ref{fig_4}(a), we show the behavior of the Fano factor for a coherent state $\ket{\alpha}$ of the field.  Similar collapses and revivals structures appear in both models. However, the photon statistics in the AJC case is sub-Poissonian during the collapse, while the JC model presents super-Poissonian statistics.  Furthermore, the Fano oscillations are also inverted, with local maxima for the AJC corresponding to local minima for the JC model (and viceversa).  The SUSY nature of both models is revealed in the Fano factor. Similarly, we plot the absolute value of the second cumulant for the field amplitude $ \langle \hat{a}^2 \rangle$ in Fig.~\ref{fig_4}(b).  The field amplitude presents an oscillatory structure, which can be approximated as $\vert\langle\hat{a}^2\rangle\vert\sim |\alpha|^2\, |\cos\pas{g(\omega_c t/|\alpha|-1)}|$.  Analogous to the Fano factor, the field amplitude reveals the SUSY nature of both models due to the antipodal oscillations involved.\\

To visualize the peculiar quantum behavior in the AJC dynamics, we select four instants of time and show the corresponding Wigner functions, defined as $W (\alpha,\alpha^*)= \frac{1}{\pi}\sum_{k=0}^\infty (-1)^k\bra{k}\hat D^\dagger(\alpha)\hat\rho\hat D(\alpha)\ket{k}$, where $\hat{D}(\alpha) = \exp\pas{\alpha \hat a^\dagger - \alpha^*\hat a}$ is the Glauber displacement operator and $\hat{\rho}$ represents the density matrix, in Figure~\ref{fig_4}(c).  At $t=0$, the Wigner function is a Gaussian centered at the value of $\alpha = 4$. The Wigner function moves anti-clockwise in a circle of radius equal to $\alpha$.  At the first maximum of FF, the Wigner function completes a rotation of $\pi$.  At the collapse, the Wigner function takes on a cat-like state structure.  Finally, we represent the complex Wigner structure at the maximum of the first FF revival.  These results provide a fingerprint of the SUSY nature of the JC and its partner, the AJC model.

\bibliography{My_Bib_SUSY_JC}
\end{document}